\documentclass[prd,aps,twocolumn,showpacs,superscriptaddress,floatfix]{revtex4-1}
\usepackage{bm}
\usepackage{amsmath}
\usepackage{txfonts}
\usepackage{graphicx}% Include figure files
\usepackage{dcolumn}% Align table columns on decimal point
\usepackage{bm}% bold math
\usepackage{color}
\usepackage{multirow}
\def\comment#1{}

\usepackage{subcaption}

\def\slashchar#1{\setbox0=\hbox{$#1$}           % set a box for #1
   \dimen0=\wd0                                 % and get its size
   \setbox1=\hbox{/} \dimen1=\wd1               % get size of /
   \ifdim\dimen0>\dimen1                        % #1 is bigger
      \rlap{\hbox to \dimen0{\hfil/\hfil}}      % so center / in box
      #1                                        % and print #1
   \else                                        % / is bigger
      \rlap{\hbox to \dimen1{\hfil$#1$\hfil}}   % so center #1
      /                                         % and print /
   \fi}                                         %

\begin{document}
%\title{Universality of Glassy Dynamics in a Quantum Framework}
%\title{Quantum mechanics motivated single parameter collapse of the viscosity of all glass formers}
%\title{A Quantum Theory of the Glass Transition Suggests Universality Amongst Glass Formers}
%\title{Universal behavior of supercooled liquids from a single parameter model: The Distributed Eigenstate Hypothesis}
\title{A phase space approach to supercooled liquids and a universal collapse of their viscosity}

\author{Nicholas B. Weingartner$^{*}$}
\affiliation{Institute of Material Science and Engineering, Washington University, St. Louis, MO 63130, U.S.A.}
\affiliation{Department of Physics, Washington University, St. Louis,
	MO 63130, U.S.A.}
\email{weingartner.n.b@wustl.edu}
\author{Chris Pueblo}
\affiliation{Institute of Material Science and Engineering, Washington University, St. Louis, MO 63130, U.S.A.}
\affiliation{Department of Physics, Washington University, St. Louis,
	MO 63130, U.S.A.}
\author{Flavio S. Nogueira}
\affiliation{Institute for Theoretical Solid State Physics, IFW Dresden, Helmholtzstr. 20, 01069 Dresden, Germany}
\affiliation{Institut f{\"u}r Theoretische Physik III, Ruhr-Universit\"at Bochum,
	Universit\"atsstra\ss e 150, DE-44801 Bochum, Germany}
\author{K. F. Kelton}
\affiliation{Institute of Material Science and Engineering, Washington University, St. Louis, MO 63130, U.S.A.}
\affiliation{Department of Physics, Washington University, St. Louis,
	MO 63130, U.S.A.}
\author{Zohar Nussinov$^{\dagger}$}
\affiliation{Institute of Material Science and Engineering, Washington University, St. Louis, MO 63130, U.S.A.}
\affiliation{Department of Physics, Washington University, St. Louis,
	MO 63130, U.S.A.}
\affiliation{Department of Condensed Matter Physics, Weizmann Institute of Science, Rehovot 76100, Israel}
\email{zohar@wuphys.wustl.edu}

\date{\today}

\begin{abstract}
A broad fundamental understanding of the mechanisms underlying the phenomenology of supercooled liquids has remained elusive, despite decades of intense exploration. When supercooled beneath its characteristic melting temperature, a liquid sees a sharp rise in its viscosity over a narrow temperature range, eventually becoming frozen on laboratory timescales. Explaining this immense increase in viscosity is one of the principle goals of condensed matter physicists. To that end, numerous theoretical frameworks have been proposed which explain and reproduce the temperature dependence of the viscosity of supercooled liquids. Each of these frameworks appears only applicable to specific classes of glassformers and each possess a number of variable parameters. Here we describe a classical framework for explaining the dynamical behavior of supercooled liquids based on statistical mechanical considerations, and possessing only a single variable parameter. This parameter varies weakly from liquid to liquid. Furthermore, as predicted by this new classical theory and its earlier quantum counterpart, we find with the aid of a small dimensionless constant that varies in size from $\sim 0.05-0.12$, a universal (16 decade) collapse of the viscosity data as a function of temperature. The collapse appears in all known types of glass forming supercooled liquids (silicates, metallic alloys, organic systems, chalcogenide, sugars, and water).
\end{abstract}


\maketitle

\section{Introduction}

Human kind has been forming and using glasses for millennia. The unique optical, thermal, and mechanical properties, as well as ease of working, that arise from the lack of long-range crystalline order in glasses \cite{gupta} has lead to their application in a diverse range of fields \cite{application1,application2,application3, application4,application5,application6}. Despite their ubiquity, a fundamental understanding of the phenomenology associated with glasses and their formation via the vitreous transition remains elusive. In order to understand the structural and mechanical behavior of glasses, we must first understand how and why they form at all. As glasses form from supercooled liquids, this means we must first understand the dynamics of supercooled liquids.
Ordinarily, when an equilibrium liquid is cooled to a temperature beneath its melting point it undergoes a first order thermodynamic transition to the ordered crystalline solid. However, if the liquid is cooled sufficiently quickly (at material dependent rate), crystallization can be bypassed, and the liquid enters a metastable (with respect to the crystal) state, and is termed ``supercooled''. The thermodynamic and kinetic properties of supercooled liquids exhibit a number of remarkable characteristics, but the most striking is arguably the behavior of the viscosity (and all associated relaxation times) \cite{1,2,3,4,5,6}. The viscosity of supercooled liquids grows by as much as 14 decades over temperature ranges as small as a few hundred Kelvin, eventually reaching a value of $10^{12} Pa*s$ at the kinetic glass ``transition" that occurs at a temperature $T_g$. Calorimetric signatures of the transition into the glassy state have also been observed at the dynamic glass transition temperature $T_{g}$\cite{yyue}. At temperatures below the glass transition temperature, $T_g$, the increasingly sluggish dynamics lead to the onset of rigidity and solid-like behavior in the liquid on observable timescales. This immense dynamical slowing occurs without any obvious structural change/ordering, and attempts to find an appropriate order parameter or growing length scale have remained inconclusive. As such, explaining the spectacular increase of the viscosity (and associated relaxation time) of supercooled liquids remains an open challenge in material science. 

Liquids which are in equilibrium at high temperatures above melting, have a viscosity which is well described by an Arrhenius function, namely 

\begin{eqnarray}
\eta(T)=\eta_0 e^{\frac{\Delta G(T)}{k_B T}},
\label{Arrhenius}
\end{eqnarray}
with $\Delta G(T)$ a (weakly) temperature dependent Gibb's free energy of activation and $k_B$ Boltzmann's constant. The simple interpretation of this form is that there exists a well-defined energy barrier (associated with bond-breaking) that can be overcome by thermal excitations. As the temperature is lowered, appropriately sized thermal fluctuations become considerably less likely and flow decreases appreciably. If this form were maintained in the supercooled liquid, there would be little mystery. However, all liquids show a degree of departure from the Arrhenius form. This degree of departure forms a continuous spectrum, and is quantified by Angell's fragility parameter \cite{bib:angell,bib:Austen}. According to this scheme, the most 
``fragile'' liquids (those with the high values of the fragility parameter) display a far more dramatic rise in the viscosity than that predicted by an Arrhenius law whereas the deviation from an Arrhenus law is far smaller in ``strong'' liquids (having a small fragility). The underlying physics of the departure from Eq. (\ref{Arrhenius}) is what we aim to explain.

Some of the first attempts to describe the non-Arrhenius character of supercooled liquid viscosity were undertaken in the 1920's by Vogel, Fulcher, Tammann, and Hesse \cite{bib:vft}. Collectively, they discovered that the functional form,
\begin{eqnarray}
\eta(T)=\eta_0e^{\frac{DT_0}{T-T_0}}
\label{VFT}
\end{eqnarray}
was able to adequately describe the viscosity of many supercooled liquids over a fair range of temperatures. In the so-called VFTH form, the parameter $D$ is related to the fragility, and $T_0$ is a material-dependent temperature at which a dynamic divergence is predicted to occur. This form initially appeared as a purely empirical form, with no rigorous theoretical support. However, over the years a number of theoretical frameworks have been proposed \cite{fv,bib:ag,bib:ag1,bib:ag2, bib:rfot,bib:mct1,bib:mct2,bib:mct3} to reproduce the VFTH form. While the VFTH form has survived for nearly a century and is widely used, it has consistently been shown to provide an overall poor fit to the viscosity of supercooled liquids of all types (classes, fragilities, bonding types, etc.) over the whole range of data. Additionally, there is no conclusive evidence for a dynamic divergence at any temperature above absolute zero \cite{Darkly}. These include tantalizing experiments that employed 20 million year old amber \cite{Darkly,critical_vft}. For these and other reasons, a plethora of other functional forms have been proposed in the last 30 years which do not contain a dynamic divergence, and which have rigorous theoretical foundations. A few of these which have been found to accurately describe the viscosity of many glass forming liquids are the KKZNT, Cohen-Grest free volume, parabolic, and MYEGA forms \cite{bib:KKZNT1,bib:KKZNT2,bib:KKZNT3,bib:CG,bib:BENK1,bib:BENK2,bib:MYEGA}. 

The aforementioned functional forms have all been shown to do an excellent job of reproducing the temperature dependence of the viscosity of a wide range of supercooled liquids. For example, the KKZNT form \cite{bib:KKZNT1,bib:KKZNT2,bib:KKZNT3} has become a favorite of some researchers in the metallic glass community and very accurately describes the behavior of metallic liquids \cite{bib:Blodgett}, while the MYEGA form \cite{bib:MYEGA} has become ubiquitous in the sillicate and oxide glass community, as it works very well for covalently-bonded non-organic liquids. The trouble with these forms, as we will show, is that despite their applicability to some liquids they do not accurately describe all types of supercooled liquids. This is made particularly striking in a review by Angell {\it et al.} \cite{bib:Austen}, in which the authors list \textit{ten} different functional forms all of which they discuss are accurate only for certain types/classes of liquids. Additionally bothersome is that most of these theories contain at least three adjustable parameters which cannot be uniquely determined by correlations with thermodynamic observables. This is true of the parabolic form \cite{bib:BENK1,bib:BENK2}, which has wide applicability to fragile glasses. It seems reasonable to expect that if any liquid can in principle be supercooled, then there should be some universal mechanism/theory that is applicable across all liquids. Further, the material dependent parameters of a given model should be related to thermodynamic observables, and not arbitrary fitting variables, while reflecting first principles.

In order to remedy the issues discussed above, we will now propose and assess a classical statistical mechanical framework to describe the viscosity of supercooled liquids. An earlier quantum rendition of our theory that mirrors and contains many of the considerations invoked in the classical approach that we discuss here first appeared in \cite{bib:DEH1} and motivated the fit and collapse that we experimentally tested and derive here classically. Within our framework, the temperature dependence of the viscosity contains only a single parameter. Such a functional dependence implies a collapse of the viscosity data. In the current work, we collapse the published viscosity data of 45 supercooled liquids onto a single scaling curve. This collapse is a central result of our work. Additional aspects of our approach (in particular, the calculation of Angell's fragility parameter and the viscosity above the melting temperature) along with further details concerning our data analysis and fits appear in \cite{bib:DEH2}. Regardless of our theoretical bias, the existence of the universal collapse of the viscosity data that we first report on here suggests (as it has in many other arenas for very different problems \cite{collapse1,collapse2,collapse3}) an underlying simplicity. Historically, the existence of a collapse in which the data from numerous systems were seen to fall on a universal curve pointed to a commonality in standard equilibrium critical phenomena \cite{collapse3}. Historically, the discovery that experimental data for various systems 
in the vicinity of their liquid to gas phase
transition can be made to collapse onto a single curve after a simple rescaling \cite{Guggenheim} 
predated current understanding of critical phenomena by many decades and 
hinted at the universality that permeates equilibrium phase transitions \cite{collapse3,gerardo}.   
We hope that the viscosity collapse that we find for all studied supercooled liquids will spur further investigation. 
In the next section, we turn to the rudiments of our classical statistical mechanics approach. 

\section{Fundamentals of the Energy Shell Distribution Approach}

The macroscopic thermodynamic and dynamical observables (such as viscosity) of a many-body system ultimately result from the average of the microscopic dynamics of the constituent atoms of the system. These microscopic dynamics are governed by the interactions between the system's constituent members, and these are encoded in the system's Hamiltonian, $H$, which is a function of the kinetic and interaction energies of the constituent atoms in the system. We can write down the \textit{\textbf{exact}} classical, many-body Hamiltonian for a supercooled liquid of any type as

\begin{eqnarray}
\label{atom}
H=\sum_i \frac{\vec{P}_i^2}{2M_i} +\sum_i \frac{\vec{p}_i^2}{2m_e}+ \sum_{i>i'} \frac{e^2}{4\pi \epsilon_0 |\vec{r}_i-\vec{r}_{i'}|}  \nonumber \\
+\sum_{ij} \frac{Z_ie^2}{4\pi \epsilon_0 |\vec{R}_j-\vec{r}_i|}+\sum_{j>j'} \frac{Z_iZ_je^2}{4\pi \epsilon_0 |\vec{R}_j-\vec{R}_{j'}|}.
\label{Hamiltonian}
\end{eqnarray}
where $Z_i$ is the atomic number, $m_e$ is the electron mass, $M_i$ is the atomic mass, $\vec{r}_i$ is the position of the i-th electron, and $\vec{R}_j$ is the position of the j-th nucleus. We consider realistic three-dimensional liquids of $N$ particles (the total number of electrons and nuclei). This Hamiltonian is intentionally general; changing the values of $Z_i$, $M_i$, and the specific form of any additional interaction potentials allows one to describe any and all \textit{specific} liquids. Although the exact Hamiltonian is given by Eq. (\ref{atom}), this precise form of the Hamiltonian will be immaterial in the very general analysis that follows. Rather, as we will explain, what matters most in our classical approach (and in its quantum analog \cite{bib:DEH1}) 
is that the equilibrium properties of this disorder free many body Hamiltonian are empirically well known. Specifically, the realization of Hamlltonian of Eq. (\ref{atom}) as it pertains to standard disorder free materials, typically exhibits equilibrium solid or liquid phase at, respectively, low or high energy densities or temperatures. 

In what briefly follows, we denote the collection of the momentum coordinates of all particles (electrons and nuclei) by $\vec{\pi}$ 
and the collection of all spatial coordinates by $\vec{x}$. To compute the dynamics of constituents of the liquid, one needs to solve Hamilton's equations,
\begin{eqnarray}
\dot{\vec{\pi}}=- \nabla_{\vec{x}} H \nonumber \\
\dot{\vec{x}}=\nabla_{\vec{\pi}} H
\label{Hamilton}
\end{eqnarray}
corresponding to the Hamiltonian in Eq.(\ref{Hamiltonian}). In general, for many-body systems such as a liquid, this leads to a set of strongly coupled, highly nonlinear, partial differential equations which are impossible to solve exactly. While methods of approximation do exist to solve the resulting equations, we need only rely on simple statistical mechanics ideas in conjunction with extensively verified experimental observations. 

\begin{figure*}
	\centering
	\includegraphics[width=1.8 \columnwidth]{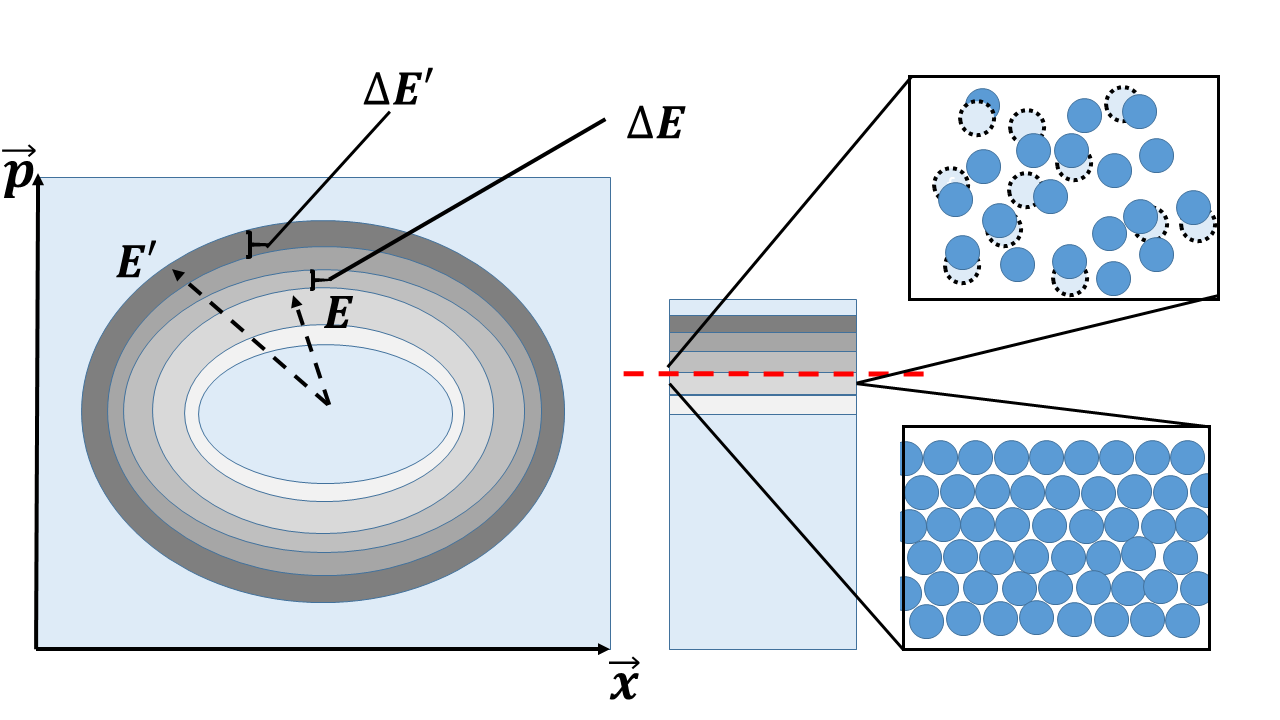}
	\caption{(Color Online.)  At left is a phase space schematic with fixed energy shells. As described in the text, the microstates change from being ``liquid like'' at high energy densities (or associated high temperatures) to being ``solid like'' at low energies (or low temperatures). On the right, we depict a cartoon of the atomic microstates both above and below the energy density associated with melting (dashed line at center).}
	\label{Shell.}
\end{figure*}

The state of a classical N-body system may be represented by a point labelled by the positions and momenta of each of the particles in 6N-dimensional phase space (a microstate $(\vec{x}, \vec{\pi})$). The time evolution of the system corresponds to a trajectory in this phase space which is governed by the system Hamiltonian and any external constraints. The system is assumed to be in any of the microstates which are allowed by the external constraints of the system (macrostate), with appropriate statistical weights set by the specific ensemble being employed. The calculation of the average values of physical observables $O$ (which correspond to the measured macroscopic values), proceeds by averaging the value of $O$ in each microstate of the allowed region of phase space. For an isolated system, the allowed phase space is given by a shell centered on the hypersurface of constant energy, $H(\vec{x},\vec{\pi})=E$, with thickness $\delta E$ set by the uncertainty in specification of the external energy. The statistical weights are constant in the allowed region and zero elsewhere, such that microcanonical averages are given by 

\begin{eqnarray}
\bar{O}(E)=\frac{1}{D(E)}\int d^{3N}{x}\int\frac{d^{3N}\pi}{(2\pi\hbar )^{3N}}~O(\vec{x},\vec{\pi})~\delta (H(\vec{x},\vec{\pi})-E), 
\nonumber\\
\label{Micro}
\end{eqnarray}
with the density of states 
\begin{equation}
\label{Eq:DOS}
D(E)=\int d^{3N}{x}\int\frac{d^{3N}\pi}{(2\pi\hbar )^{3N}}~\delta(H(\vec{x},\vec{\pi})-E).
\end{equation}
When the system is coupled to an external heat bath, all energies are in principle attainable by the system, and the infinitesimally thick shell ($\delta$-peaked) of allowed phase space (Fig. (\ref{Shell.})) may become smeared and overlap. This leads to averages of the form
\begin{eqnarray}
\bar{O}=\int dE'~\bar{O}_{m}(E')~\rho(E').
\label{Dist}
\end{eqnarray}
Here, $\bar{O}_m(E')$ is the microcanonical average at energy $E'$, and $\rho(E')$ is a (normalized) probability distribution in phase space which is not a $\delta$-function. 
In standard equilibrated systems (such as those corresponding to the disorder free Hamiltonian of Eq. (\ref{atom}) that describes equilibrated solids and liquids), the ensemble average of Eq. (\ref{Dist}) is equal to the long time average of $O$ (which we denote by $O_{\infty}$) as it evolves according to Eqs. (\ref{Hamilton}). Empirically, as we remarked earlier, at high enough temperatures or energy densities, the system of Eq. (\ref{atom}) is a fluid while at temperatures or energy densities below that of freezing the system is an equilibrium solids. Thus, for {\it any observable} $O$, the microstate average of Eq. (\ref{Micro}) will change character from featuring equilibrium fluid like features at high energies to solid like behaviors at low energies. When latent heat appears at the equilibrium melting transition (as it nearly always does), there will be intermediate states displaying mixed fluid and solid features. It follows that, when averaged over energy shells in phase space, the microstates themselves change their character across the equilibrium phase transitions. Fig. \ref{Shell.} portrays the above simple conclusion.

Since 
systems in equilibrium, with a well defined temperature, have a canonical partition function,
\begin{equation}
Z=\int_{-\infty}^\infty dE~D(E)~e^{-\frac{E}{k_BT}}.
\end{equation} 
In this case, $\rho(E')$ corresponds o the Gibbs distribution, namely, 
\begin{eqnarray}
\rho(E')=\frac{D(E') e^{-\frac{E'}{k_B T}}}{Z}.
\label{Gibbs}
\end{eqnarray}
If the system is now cooled quasistatically, equilibrium will be maintained, and the distribution will remain canonical at progressively lower temperatures. This is, in part, guaranteed by Liouville's theorem, which states that the phase space volume along trajectories in phase space is preserved for Hamiltonian systems. This means that as the system is cooled slowly enough, trajectories will neither bunch nor diverge and will map in a ``1-to-1'' fashion to the newly allowed region of phase space, and the distribution function will adjust accordingly. If instead of slow quasistatic cooling, we rapidly quench the system, it will cease to be in equilibrium and its dynamics will no longer be Hamiltonian. The now dissipative system will violate Liouville's theorem: the trajectories from nearby points in phase space can diverge, and the phase space volume may swell. The initial shape of the initial energy shells will be deformed due to supercooling deform. This deformation is central to our description of the supercooling process. Due to this non-adiabatic evolution, the Gibbs distribution will no longer be the exactly correct distribution describing the distribution in phase space. If we allow the system to maintain metastable equilibrium then the canonical ensemble is still roughly obeyed. However, in this case, different regions of the initial phase space will map to areas with different effective canonical distribution functions, i.e., with different effective temperatures. This idea, which is seemingly reinforced by the appearance of dynamical heterogeneities \cite{bib:DEH1,DH1,DH2,DH3,DH4,DH5,DH6} and other phenomena implies that the overall system will sample a range of effective global temperatures (necessitated by the apparent spatial distribution of local effective temperatures) consistent with the externally imposed temperature, $T$. This distribution of effective temperatures forms the nub of our ``Energy Shell Distribution Theory'' (ESDT).  

With the system now sampling a smeared out distribution of effective temperatures, the phase space probability distribution for the averages of Eq. (\ref{Dist}), will now involve a conditional probability density $\rho(E|T')$ for the energy given a specific temperature, namely
\begin{eqnarray}
\rho(E)=\int dT'~\rho(E|T')\rho(T').
\label{Conditional}
\end{eqnarray}
Here, $\rho(T')$ is the probability distribution of effective temperatures $T'$. 
As the system is in a metastable equilibrium, the probability density for a given $E$ at a temperature $T'$ will still reasonably be described by the Gibbs distribution of Eq. (\ref{Gibbs}).
Similar to Eq. (\ref{Dist}), the long time average of $O$ for a general distribution $\rho$ including that associated 
with the supercooled liquid ($sc$) reads \cite{bib:DEH1}
\begin{eqnarray}
\bar{O}_{\infty,sc} %=\int dT'\rho(T')~\frac{\int dE \rho(E|T')~O(E)}{Z(T')} \nonumber \\
%=\int dT' \rho(T')~ \frac{\int dE~e^{-\frac{E}{k_B T'}}~O(E)}{Z(T')} \nonumber \\
=\int dT' \rho(T')~ \tilde{O}_{can}(T').
\label{ESDTA} 
\end{eqnarray}
Here, $\tilde{O}_{can}(T')$ is the canonical, equilibrium value of the observable ${\cal{O}}$ at a temperature $T'$. We see, then, that supercooling acts to drive the system into a metastable equilibrium which leads to the system sampling a range of equilibrium value averages over a narrow, but finite distribution of effective temperatures. The initial ``shock" to the system of supercooling causes microscopic effects which broaden the distribution. By virtue of being out of equilibrium, the distribution $\rho$ must have a finite standard deviation. This is so as otherwise the system would be described by a unique uniform effective temperature and be describable by the equilibrium canonical ensemble. However, since the supercooled liquid is out of equilibrium, the standard deviation $\sigma$ associated with the distribution $\rho$ cannot vanish \cite{bib:DEH1}. When thermodynamic equilibrium is restored at a uniform global temperature $T$, the distribution $\rho(T')$ becomes a delta function ($\delta (T-T')$) implying an equilibrium Boltzmann distribution (and ensuing equilibrium expectation values for all observables). 

With the statistical mechanics ideas in place, we now invoke these to calculate the values of observables of interest. One method of measuring the viscosity of a liquid is by measuring the terminal velocity of a sphere dropped into the liquid. In this case, the viscosity is inversely proportional ($\eta \propto 1/v_{\infty}$) to the terminal velocity of the sphere. The terminal velocity is a macroscopic property of the system, and therefore can be calculated in our statistical mechanical framework. Setting the observable $O$ to be the vertical velocity of the dropped sphere, $O=v_{z}$ \cite{bib:DEH1}, the observed terminal velocity becomes
\begin{eqnarray}
\bar{v}_{\infty,sc}=\int dT' \rho(T')~\tilde{v}_{\infty,can}(T').
\label{vel}
\end{eqnarray}
Thus, the viscosity will be given by
\begin{eqnarray}
\eta=\frac{A}{\int dT' \rho(T')~\tilde{v}_{\infty,can}(T')},
\label{Visc}
\end{eqnarray}
with $A$ a constant. As is well known, for an equilibrium system, there exists a cutoff temperature, $T_c$, below which the the terminal velocity must vanish (since the system is completely solid
and no long time flow occurs). Thus, in the equilibrium canonical ensemble, only averages of the terminal velocity at temperatures above this cutoff may contribute to the integral in Eq. (\ref{Visc}) leading to 
\begin{eqnarray}
\eta=\frac{A}{\int_{T_c}^{\infty} dT' \rho(T')~\tilde{v}_{\infty,can}(T')}.
\label{Visc1}
\end{eqnarray}
If we further assume that the distribution $\rho$ is sufficiently narrowly peaked (as will be verified in the next section and seen from the numerical values of our dimensionless fit parameter) such that the distribution has minimal ``leakage" into effective temperatures $T'$ above $T_c$, when the measured global temperature $T<T_c$, then the value of $v_{\infty,can}$ will change very little over the region of appreciable weight. Therefore, we can reasonably replace $v_{\infty,can}(T')$ with $v_{\infty,can}(T_c)$. Thus the viscosity of the supercooled liquid is 
\begin{eqnarray}
\eta=\frac{\eta(T_c)}{\int_{T_c}^{\infty} dT' \rho(T')}.
\label{Visc2}
\end{eqnarray}
In order to use this expression to make concrete predictions of the viscosity, we must know what functional form to use for $\rho(T')$. All that is known about the distribution is that it is peaked about the external temperature, $T$, must be normalized, and that it has a small yet finite width. In the absence of additional constraints, the appropriate distribution $\rho$ for the supercooled liquid may be ascertained \cite{bib:DEH1} by maximizing the Shannon entropy $H_I= -\int\rho(T') \log_{2}[\rho(T')]~ dT'$. As is well known, maximizing the Shannon entropy with the constraints of normalization and finite variance leads to a Gaussian distribution. Therefore, the most probable distribution of effective temperatures is
\begin{eqnarray}
\rho(T')=\frac{1}{\sqrt{2\pi}\sigma(T')}e^{-\frac{(T'-T)^2}{2\sigma(T')^2}}
\label{Gauss}
\end{eqnarray}
where $\sigma(T')$ represents the spread of the distribution and $T$ is the external temperature. 
Inserting the Gaussian distribution of Eq.(\ref{Gauss}) into Eq.(\ref{Visc2}), we find that the viscosity 
\begin{eqnarray}
\eta(T)=\frac{\eta(T_c)}{erfc\left[\frac{T_c-T}{\sqrt{2}~\sigma(T)}\right]}.
\label{Visc3}
\end{eqnarray}
In what follows, we make two conjectures to complete the form of the viscosity, one involving the cut-off temperature, and one involving the spread of the Gaussian. 

\subsection{The cutoff temperature}

\begin{figure*}
	\centering
	\includegraphics[width=1.6 \columnwidth]{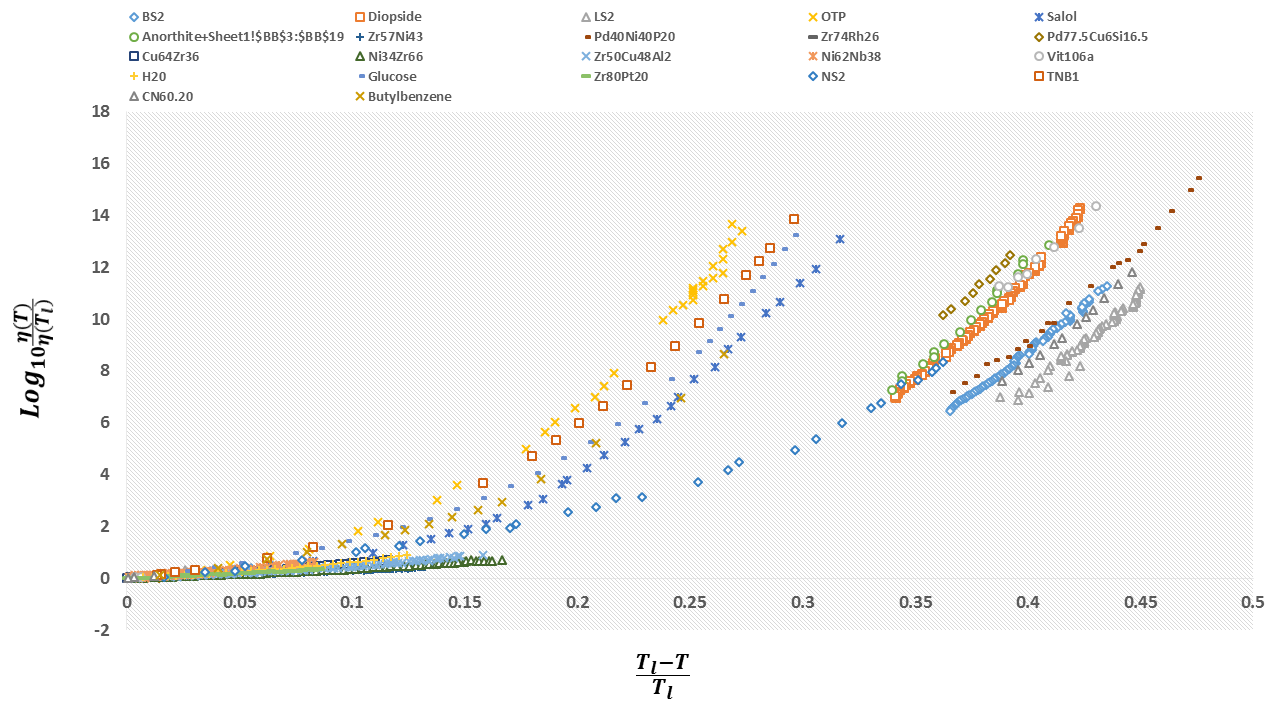}
	\caption{(Color Online.) The viscosity, $\eta(T)$, scaled by its value at the melting (or liquidus) temperature $\eta(T_{l})$ plotted as a function of the ``reduced temperature''
	$\frac{T_{l}-T}{T_{l}}$. When represented this way, a spectrum of behaviors appears, with most glassformers seeming to fall within different `families' corresponding to fragility classes as defined by experimental values. }
	\label{Fragility.}
\end{figure*}

\begin{figure*}
	\centering
	\includegraphics[width=2 \columnwidth, height=.7 \textheight, keepaspectratio]{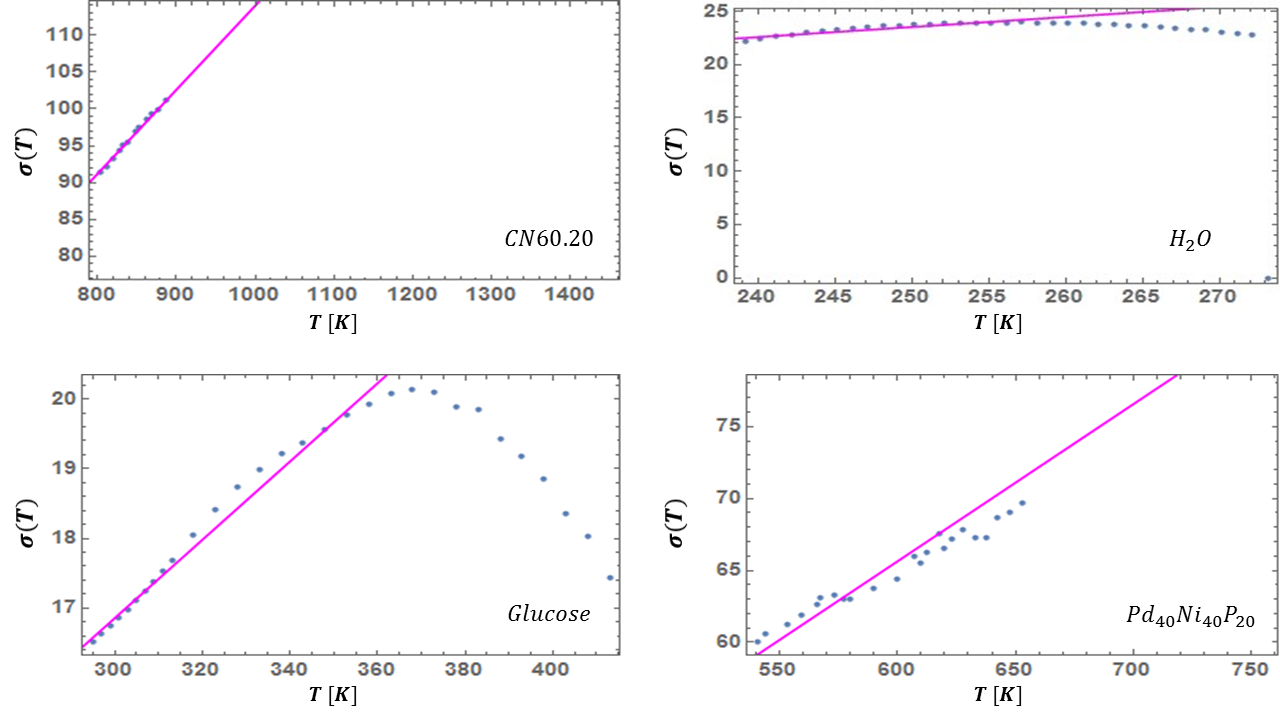}
	\caption{(Color Online). The standard deviation $\sigma(T)$ of the probability distribution of Eq. (\ref{Gauss}) as inferred by fitting the experimentally measured viscosity to Eq. (\ref{ESDT}). In most cases that we examined, the approximate linearity relation of Eq. (\ref{sigma}) holds reasonably well far enough below the liquidus temperature. Here, we also show
	two well known exceptional liquids: water and glucose. These fluids display anomalies that have been ascribed to putative liquid-liquid transitions, e.g., \cite{water1,water2,water3,water4,water5,glucose1,glucose2}. The crossover of $\sigma$ at high temperature 
	and the one that we similarly found in supercooled salol \cite{bib:DEH2} may be a signature of these putative transitions.
	Indeed, in salol the crossover temperature at which $\sigma(T)$ deviates from its low temperature linearity  \cite{bib:DEH2} coincides with earlier experimentally suggested liquid-liquid transition temperature \cite{salol}.}  
	\label{linearT}
\end{figure*}

In choosing a value for the cut-off temperature, $T_c$, we rely on experimental observations. In pure systems, at the melting temperature, the equilibrium system undergoes a first order phase transition from the liquid to the ordered crystalline solid state. At this temperature, in equilibrium, the values of thermodynamic observables transition from their liquid-like values to their solid-like ones. In a perfect crystal (an idealization never realized), the viscosity is infinite \cite{ddd}, and hence the terminal velocity will be zero at temperatures beneath melting.

The idea of linking the glass transition to melting goes back decades and it is easy to understand why \cite{Melt,Melt2,Melt3,Melt4}. By definition, supercooled liquids are formed by avoiding crystallization at the melting transition, therefore the melting temperature implicitly determines at which temperatures a supercooled liquid exists at all. Additionally, the melting transition occurs at a sharp transition temperature, making it a somewhat less arbitrary reference point than the kinetically defined glass transition temperature. Kauzmann, in his seminal paper \cite{Kauzmann}, was one of the first to propose an empirical link between the glass transition and melting. He observed that for all the liquids he studied, on average the glass transition and melting temperatures were related by $T_g \approx \frac{2}{3} T_m$. In the intervening years, a number of researchers have found that this relationship holds, on average, for various types of supercooled liquids/glasses \cite{231,232}. However, deviations from this empirical rule have also been observed for decades. Similar to the argument above, Turnbull reasoned that because nucleation and growth of the crystalline phase became thermodynamically possible at the melting temperature, glass formability may be linked to the gap between the melting temperature and glass transition temperature. He observed that glass formability in metallic liquids could roughly be quantified by what he defined as the reduced glass transition temperature, $T_{rg}=\frac{T_g}{T_m}$ \cite{Turnbull,GFA1,Melt4}, where the best glass formers had $T_{rg}\approx \frac{2}{3}$. However, as metallic liquids display a range of glass formability, so to does the reduced glass transition temperature. Therefore, in metallic liquids at least, the 2/3 rule does not always hold. Building on the observation of Kauzmann and Turnbull, it appears reasonable to investigate further what links exist between melting and the glass transition. What these empirical relationships fail to do, however, is provide a consistent framework for understanding the dynamics of supercooled liquids and making predictions about the phenomenology based on melting. This is made vivid by examining a simple scaling of the viscosities of several liquids by values associated with melting. In Fig. (\ref{Fragility.}) we plot the logarithm of the viscosity scaled by its value at melting (or, more precisely, its liquidus temperature, as will be discussed below) versus the melting-scaled inverse temperature. As the figure demonstrates, a universal description of the viscosity does not immediately emerge by simply using the melting temperature, however ``fragility bands" appear, providing more evidence for the link between $T_g$ and $T_m$. This suggests that an ``ingredient" is missing. It is our goal to combine the above ideas with our simple statistical mechanical treatment, to ultimately arrive at a complete, predictive theory of supercooled liquids.

\begin{table*}
	\centering
	\caption{Values of Relevant Parameters for all liquids studied}
	\resizebox*{!}{\textheight}
	{\begin{tabular}{*{4}{@{\hskip 0.4in}c@{\hskip 0.4in}}} 
			\toprule
			\emph{Composition} \quad & \emph{}$\bar{A}$ & \emph{$T_l$ [K]} & \emph{$\eta(T_l)$ [\mbox{Pa*s}]} \\
			\colrule
			BS2 & 0.157129 & 1699 & 5.57   \\
			Diopside & 0.134328 & 1664 & 1.50     \\
			LS2 & 0.170384 & 1307 & 22.19     \\
			OTP & 0.069685 & 329.35 & 0.029    \\
			Salol & 0.087192  & 315 & 0.008   \\
			Anorthite & 0.131345 & 1823 & 39.81  \\
			Zr$_{57}$Ni$_{43}$ & 0.234171 & 1450 & 0.015    \\
			Pd$_{40}$Ni$_{40}$P$_{20}$ & 0.154701 & 1030 & 0.03019  \\
			Zr$_{74}$Rh$_{26}$ & 0.187851 & 1350 & 0.036    \\
			Pd$_{77.5}$Cu$_6$Si$_{16.5}$ &  0.124879 & 1058 & 0.044     \\
			Albite & 0.103344 & 1393 & 24154952.8 \\
			Cu$_{64}$Zr$_{36}$ &  0.142960 & 1230 & 0.021      \\
			Ni$_{34}$Zr$_{66}$ & 0.209359 & 1283 & 0.0269     \\
			Zr$_{50}$Cu$_{48}$Al$_{2}$ &0.167270 & 1220 & 0.0233     \\
			Ni$_{62}$Nb$_{38}$ & 0.109488  & 1483 & 0.042      \\
			Vit106a &0.133724  & 1125 & 0.131      \\
			Cu$_{55}$Zr$_{45}$ & 0.144521 & 1193 & 0.0266     \\
			H$_2$O & 0.133069 & 273.15 & 0.00179  \\
			Glucose & 0.079455 & 419 & 0.53       \\
			Glycerol &  0.108834 & 290.9 & 1.995    \\
			Ti$_{40}$Zr$_{10}$Cu$_{30}$Pd$_{20}$ &0.185389  & 1279.226 & 0.0165   \\
			Zr$_{70}$Pd$_{30}$ &  0.21073 & 1350.789 & 0.0228  \\
			Zr$_{80}$Pt$_{20}$ & 0.169362 & 1363.789 & 0.0480    \\
			NS2 & 0.134626 & 1147 & 992.274 \\
			Cu$_{60}$Zr$_{20}$Ti$_{20}$ & 0.103380 & 1125.409 & 0.0452    \\
			Cu$_{69}$Zr$_{31}$         & 0.157480  & 1313     & 0.0115    \\
			Cu$_{46}$Zr$_{54}$         &  0.156955 & 1198     & 0.02044 \\
			Ni$_{24}$Zr$_{76}$         & 0.244979 & 1233     & 0.02625 \\
			Cu$_{50}$Zr$_{42.5}$Ti$_{7.5}$  & 0.148249 & 1152     & 0.0268     \\
			D Fructose       &  0.050124 & 418      & 7.3155 \\
			TNB1             &  0.07567 & 472      & 0.0399 \\
			Selenium         & 0.130819  & 494      & 2.951    \\
			CN60.40          & 0.149085  & 1170     & 186.208   \\
			CN60.20          & 0.161171  & 1450     & 12.5887 \\
			Pd$_{82}$Si$_{18}$         &  0.137623 & 1071     & 0.03615 \\
			Cu$_{50}$Zr$_{45}$Al$_{5}$     & 0.118631   & 1173     & 0.0379    \\
			Ti$_{40}$Zr$_{10}$Cu$_{36}$Pd$_{14}$ &  0.137753 & 1185     & 0.0256     \\
			Cu$_{50}$Zr$_{50}$       & 0.166699 & 1226     & 0.02162    \\
			Isopropylbenzene &   0.073845 & 177      & 0.086      \\
			ButylBenzene     & 0.085066  & 185      & 0.0992     \\
			Cu$_{58}$Zr$_{42}$         &  0.131969 & 1199     & 0.02526    \\
			Vit 1 & 0.111185 & 937 & 36.598 \\
			Trehalose & 0.071056  & 473 & 2.718 \\
			Sec-Butylbenzene & 0.080088 & 190.3 & 0.071 \\
			SiO$_2$ & 0.090948 & 1873 & 1.196x$10^{8}$ \\
			\botrule      
		\end{tabular}}
		\label{Params}
	\end{table*}

In light of the above arguments we will identify the cut-off temperature $T_c$ with the melting temperature, $T_m$. There is an intrinsic difficulty in doing this, however, which must be addressed. Only certain non-monatomic liquids possess a single ``melting" temperature. In reality, most liquids have a ``melting range" associated with the temperatures between the solidus temperature $T_s$ and the liquidus temperature $T_l$. Additionally, either associated with these temperatures, or the pure-system melting temperature, $T_m$, there will be a range of energies corresponding to the latent heats/enthalpies of formation. Therefore, regardless of which temperature we choose to represent ``melting", there will be corrections necessary to account for the melting range. Additionally, many sillicate systems are polymorphic in the crystalline solid state, meaning that at various temperatures below the liquidus, the crystal transitions between different thermodynamically stable crystalline configurations. These polymorphs and their associated temperatures can have a very large impact on the thermodynamic properties of the system, with minimal apparent impact on the dynamical properties. One may obtain bounds on the viscosity by setting the cutoff or melting temperature in Eq. (\ref{vel}) to mean the liquidus temperature \cite{bib:DEH1}. If no long time flow appears in this intermediate temperature regime (i.e., if the terminal velocity of Eq. (\ref{vel}) vanishes), then this substitution of $T_{c} = T_{l}$ in Eqs. (\ref{Visc2},\ref{Visc3}) will be precise. Thus, because solid-like characteristics will \textit{first appear} at the liquidus temperature, we will take it to define the melting temperature at which point there is a change in the \textit{equilibrium} dynamics of the system. 
This argument can be further understood in the context of the Lindemann criterion. In Lindemann's model, the break down of solidity and onset of flow at the melting temperature is due to the average amplitude of vibration becoming an appreciable fraction of the lattice length ($\approx 10 \%$). At the temperature where this occurs, the lattice destabilizes and constituents become liquid like. The average amplitude of vibration is proportional to the kinetic energy, so this can be seen as the average kinetic energy of the constituents becoming enough to globally overcome the average interatomic bond strength. Observations suggest that a Lindemann-like model also holds for the devitrification of glasses \cite{Devit}. Therefore, viewing this from the perspective of cooling, at the melting (liquidus) temperature, the ``stickiness" of the interaction forces/energy first starts to dominate the kinetic energy, and the constituents begin to more strongly interact
Inserting the liquidus temperature, $T_l$, into Eq.(\ref{Visc3}), we obtain that $\eta(T)=\frac{\eta(T_l)}{erfc\left[\frac{T_l-T}{\sqrt{2}~\sigma(T)}\right]}$
We next motivate a specific functional form for the distribution $\sigma(T)$.

\subsection{The width of the distribution}

The spread in effective temperatures, $T'$, at a given external temperature, $T$, is quantified by $\sigma(T)$. This spread (related to the variance by a simple square root) is the fundamental variable in the ESDT, as it is caused by, and leads to, the metastable, non-canonical spread in temperatures/energies
Much like the exact distribution of temperatures which it governs, we do not know a priori what its functional form should be. However, there are a number of physical constraints that will ultimately motivate its exact form. As the system is cooled, the peak of the distribution (Eq. (\ref{Gauss})) shifts downward as it is centered on the external temperature, $T$. The tails, and not the peak, though, control how likely a macroscopic flow event will be. In order that the flow continue to decrease rapidly as the temperature is lowered, the width of the distribution will also have to shrink to ``pull" the tail out of sampling the flowing states. Additionally, as the system approaches absolute zero, the third law of thermodynamics will require that the spread in energies (and hence effective temperatures) vanish, such that $\sigma(T)$ must be a decreasing function of temperature. It is also readily obvious, that the only natural energy scale for the metastable supercooled liquid is set by the external temperature. Therefore, it is reasonable to assume that $\sigma(T) \propto T$. We these simple facts in mind, we assert that
\begin{eqnarray}
\sigma(T)=\bar{A} T,
\label{sigma}
\end{eqnarray}
where $\bar{A}$ is a small, dimensionless, material-dependent parameter. That is, the width $\sigma(T)$ is set by the natural energy (temperature) scale of the system. Additional analysis is provided in \cite{bib:DEH1}. To confirm the validity of this approximation, we can invert Eq.(\ref{Visc3}) solving for the spread, $\sigma(T)$, and examine it for experimental viscosity data. Across the different examined liquids, we found this to hold relatively well. In some materials, there are deviations from linearity in the vicinity of their respective solidus and/or liquidus temperatures. This is illustrated in Fig. \ref{linearT}. As seen therein, in both glucose and supercooled water (Fig. (\ref{linearT})), $\sigma(T)$ exhibits such a crossover. We found an analogous trend in supercooled salol where the crossover temperature associated with $\sigma(T)$  \cite{bib:DEH2} coincided with the earlier reported putative liquid-liquid transition temperature in this system \cite{salol}. Similarly, supercooled water and glucose display anomalies that have been ascribed to a liquid-liquid transition \cite{water1,water2,water3,water4,water5,glucose1,glucose2}. Taken together, these data suggest that, if and when present, fragile to strong crossovers or liquid-liquid phase transitions \cite{liq-liq} may be associated with deviations in $\sigma(T)$. This will be critically addressed in depth in a follow-up paper where we will further extensively demonstrate that $\bar{A}$ strongly correlates with various thermodynamic parameters and may allow for the prediction of low temperature viscosity from purely high temperature measurements. 

For the time being, we stress that $\bar{A}$ constitutes the {\it only adjustable parameter} in this framework. When combining this with Eq.(\ref{Visc3}), we now arrive, via classical phase space considerations, at our principal result 
for the viscosity \cite{bib:DEH1},
\begin{eqnarray}
\eta(T)=\frac{\eta(T_l)}{erfc\left[\frac{T_l-T}{\sqrt{2}~\bar{A}~T}\right]}.
\label{ESDT}
\end{eqnarray}
It is immediately clear from an examination of Eq. (\ref{ESDT}) that our model does not possess a dynamical singularity. In fact, if one were to calculate the entropy difference between the supercooled liquid and equilibrium crystalline solid, it would be apparent that the excess entropy could only vanish at a point where the temperature distribution becomes a delta function. When this occurs, however, the system will, by definition have returned to equilibrium. Therefore, our approach makes it plain that there cannot be a finite temperature singularity, and that the above excess entropy can only vanish if the system regains equilibrium. The function of Eq.(\ref{ESDT}) relies only on measurable quantities associated with the liquidus and a single parameter. While the specific form of the above equation is only applicable beneath the liquidus temperature, in \cite{bib:DEH2} we derived an extension to all temperatures above the liquidus, completing the theoretical model. 

\begin{figure*}
	\centering
	\includegraphics[width=2 \columnwidth, height=.7 \textheight, keepaspectratio]{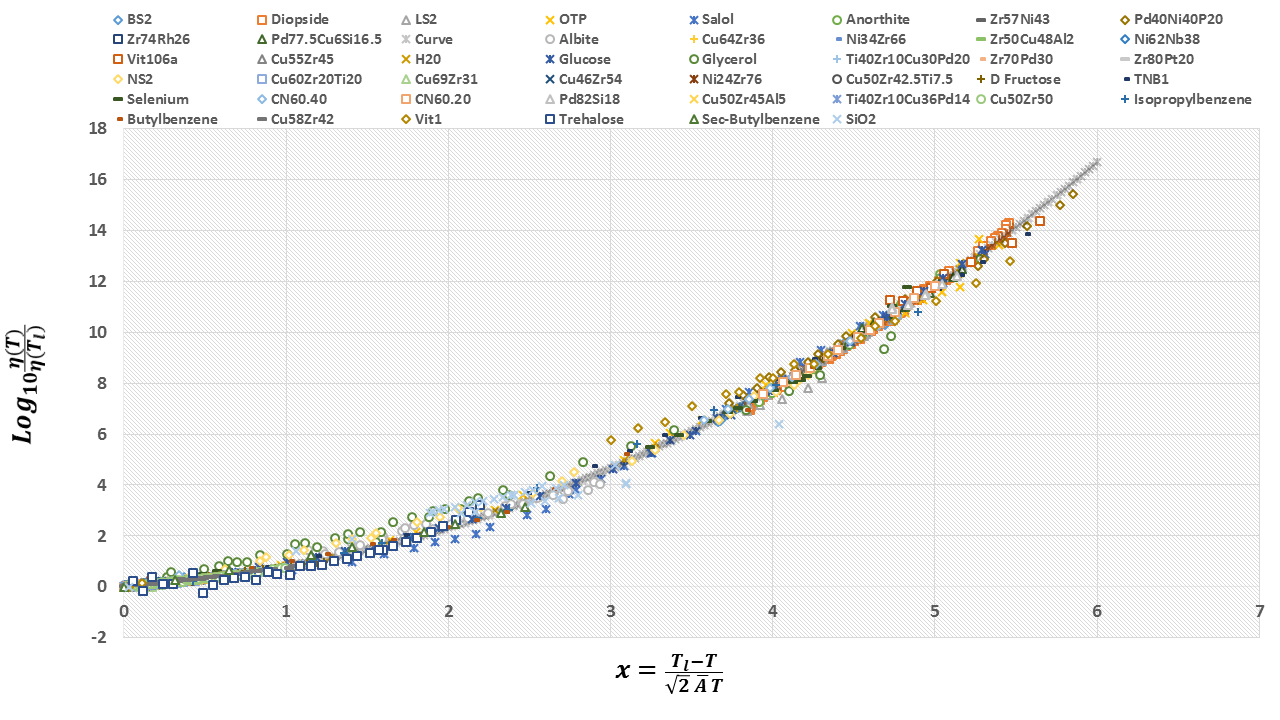}
	\caption{(Color Online.) The viscosity data scaled by its value at the liquidus temperature, $\eta(T_{l})$, versus $x$, as defined in the figure. The viscosity data of 45 liquids from numerous classes/bonding types (sillicate, metallic, organic) and kinetic fragilities collapse onto a unique curve, suggestive of universality amongst all types of glassforming liquids.  Note the exceptional agreement over 16 decades. The deviations of glycerol and SiO$_{2}$ are discussed in \cite{bib:DEH2}. The pertinent liquidus temperature $T_{l}$ and the viscosity at $T_{l}$ and our single dimensionless parameter associated with all  fluids is provided in Table \ref{Params}. The continuous underlying ``curve'' (seen at the high viscosity end where fewer viscosity data are available) is that predicted by Eq. (\ref{ESDT}).}
	\label{Collapse.}
\end{figure*}

A corollary of Eq.(\ref{ESDT}) is that the viscosity data from all supercooled liquids may be made to collapse onto one master curve.
That is, for each fluid, the ratio of the viscosity at temperature $T \le T_{l}$ to its viscosity at the liquidus temperature, $(\eta(T)/\eta(T_{l}))$, is a trivial function of the quotient $(T_l-T)/(\bar{A}~T)$ with $\bar{A}$ being
the single dimensionless parameter that is material dependent. We tested this prediction in Fig. (\ref{Collapse.}) and found it is indeed be satisfied. 
Although the value of $\bar{A}$ does not significantly change across all of the liquids that we examined 
(see Table \ref{Params}), its variations are nevertheless important. In particular, it can be demonstrated that the fragility parameter is a function of both $\bar{A}$ and 
the reduced glass transition temperature $T_{rg}$ (that are set, in our theory, by the values of the melting temperature and $\bar{A}$ themselves) \cite{bib:DEH2}. Thus, albeit being small in size, the changes in the values of $\bar{A}$ in their relatively narrow range (along with the values of $T_{rg}$) differentiate strong fluids from fragile ones. This is clearly seen in Fig. \ref{Fragility.}; if the dependence on $\bar{A}$ between different glass formers were weak the viscosity data in Fig. \ref{Fragility.} would have collapsed onto a single curve. The contrast between Fig. \ref{Fragility.} and Fig. \ref{Collapse.} (in which $\bar{A}$ was, for each liquid, set to the value given by Table \ref{Params}) highlights the importance of the deviations in the parameter $\bar{A}$ (the ``missing ingredient'' the we alluded to above) from one fluid to another.

\section{Methods: A test of the predicted viscosity and a data collapse}

	 \begin{figure*}
	\centering
	\includegraphics[width=1.8 \columnwidth]{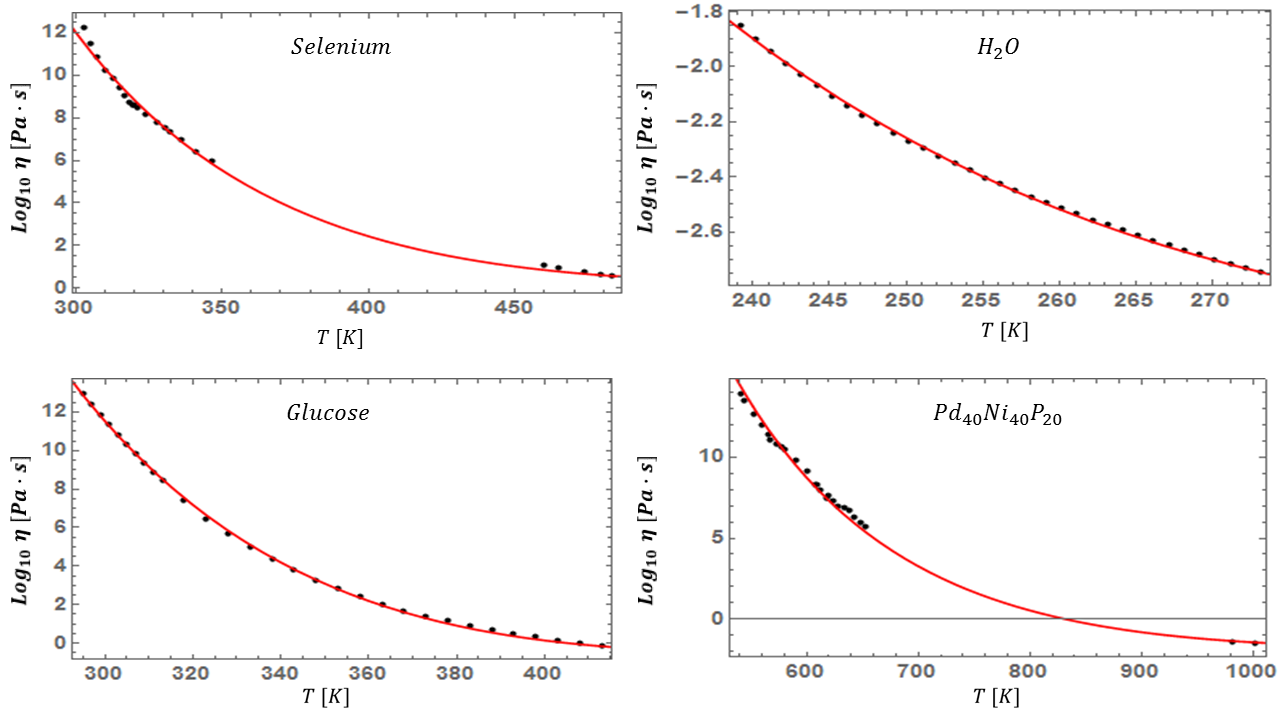}
	\caption{(Color Online.) Fits the viscosity of various supercooled fluids (including water) with Eq. (\ref{ESDT}).}
	\label{Data:OTP}
\end{figure*}

\begin{figure*}
	\centering
\includegraphics[width=1\linewidth,height=0.3 \textheight]{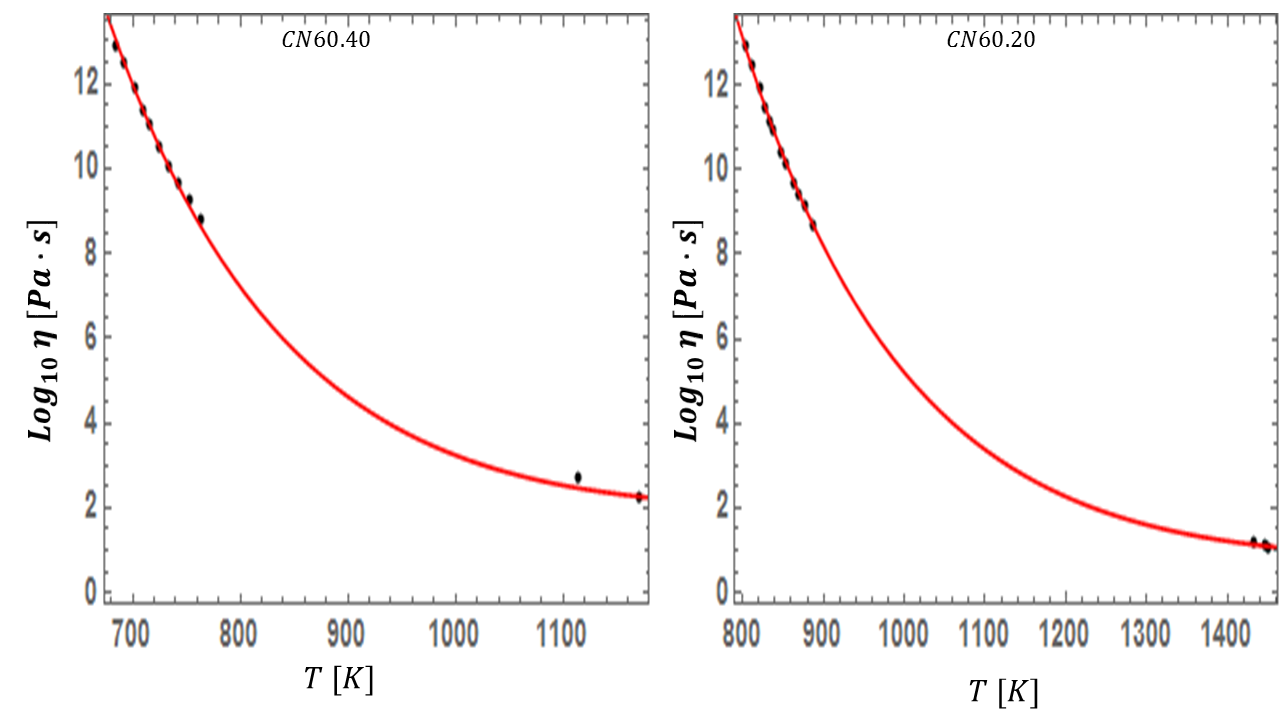}
	\caption{(Color Online.) The fit of Eq. (\ref{ESDT}) is tested for CN60.40 and CN60.20, two silicate systems with slightly different molar compositions.}
	\label{Data:CN}
\end{figure*}

\begin{figure*}
	\centering
	\includegraphics[width=1.8 \columnwidth]{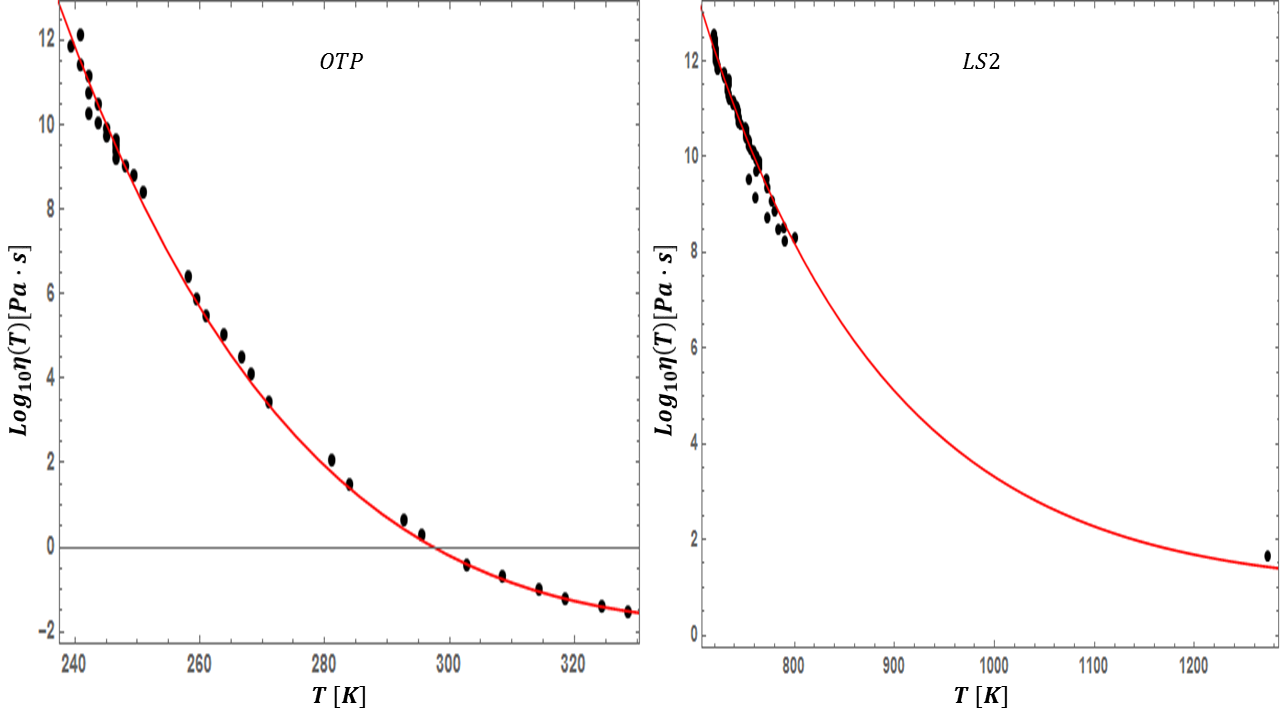}
	\caption{(Color Online.) Our viscosity fit of Eq. (\ref{ESDT}) is applied to a very fragile organic glass former (OTP) and very strong silicate glass former (LS2).}
	\label{OTP}
\end{figure*}

With the theoretical prediction of Eq.(\ref{ESDT}) in hand, we now turn our attention to assessing the accuracy of the model. We examined diverse liquids belonging to both the strong and fragile classifications, and spanning all liquid types: sillicate, oxide, metallic, organic, chalcogenide, sugars, and even supercooled water. We used standard nonlinear fitting techniques to extract the optimal value of $\bar{A}$ for each liquid. Surprisingly, the values of $\bar{A}$ fall within a small range of 0.24-0.4, while simultaneously providing a visually accurate fit to the data of all studied systems. In Table \ref{Params} we list the values of $\bar{A}$ for all studied liquids and their corresponding liquidus temperatures. In Figs. (\ref{Data:OTP}, \ref{Data:CN}, \ref{OTP}), 
 we present the viscosity data of a representative sample of the supercooled liquids studied. In the figures, the solid line represents the fit of the ESDT viscosity function to the data. Visual examination of the quality of fit suggests a high degree of accuracy. To make this objective, and quantitatively rigorous, we performed a detailed statistical analysis of the data of all liquids. In Table \ref{Stats} we present the results of the analysis, featuring the computed values for the sum of squared errors (SSE), reduced $\chi^2$, and $R^2$ statistics for each liquid. The highest $\chi^2$ values calculated (with correspondingly low $R^2$ values) correspond to glycerol and SiO$_2$, which appear to have anomalous behavior \cite{bib:DEH2} and will be further addressed in detail in a follow-up paper (the melting range and bimodality likely play a role). Outside of these two liquids, the highest value of $\chi_2$ is 0.7, and the lowest $R^2$ is 0.87. These results provide an objective validation of the ESDT model performance. From the combination of visual inspection and statistical analysis, it is within reason to conclude that the ESDT form for the viscosity is able to reproduce/describe the viscosity of all studied types of liquids to objectively high degrees of statistical accuracy. 

Whether or not the dynamics of supercooled liquids are universal has been debated for some time. We demonstrated that the ESDT viscosity form appears to fit the viscosity data of all types of supercooled liquids, thereby providing the ``missing ingredient" that prevented a universal description of liquids based on melting. If the ESDT form is to be a complete picture for all liquids, then it should allow for a universal scaling of the viscosity of supercooled liquids. For that reason, we plot the logarithm of the viscosity of all studied liquids scaled by its value at the liquidus, but this time versus the argument of the complementary error function. The results of this scaling are presented in Figure \ref{Collapse.}. It is immediately clear that this scaling collapses the viscosity data of all liquid types onto a single curve. More significantly, the collapse holds over 16 decades, and for all classes/types of liquids. It should be pointed out that while this scaling arose as a consequence of the ESDT framework, even if the theoretical foundations do not hold, this scaling can always be done. While the analysis of considerably more liquids is ultimately required, this stunning result suggests that there is perhaps an underlying universality to the dynamics of all supercooled liquids. 

For completeness, it must be pointed out that all liquids tested in this work undergo congruent melting and can therefore be adequately described by their liquidus temperature. There are numerous liquids, however, that undergo incongruent melting, and a small molar addition of some material can drastically change the liquidus temperature without appreciably effecting the viscosity. This is because the liquidus is where the small crystalline clusters associated with the addition will first appear, but in small enough concentrations, they cannot impact the dynamical character of the liquid. This presents a difficulty for using the liquidus as the scaling temperature for all liquids. The impact of this will be investigated in a further work, but suggests that for these ``pathological" liquids, a description in terms of the solidus or associated temperatures may be more appropriate.

\begin{table*}
	\centering
	\caption{Statistical measures of the goodness of the fit}
	\label{Stats}
	\resizebox*{!}{\textheight}
	{\begin{tabular}{*{4}{@{\hskip 0.4in}c@{\hskip 0.4in}}}
			\toprule
			\emph{Composition} & \emph{$(SSE)$} & \emph{$\chi_{red}^2$} & \emph{$R^2$} \\
			\colrule
			{OTP} & 10.617 & 0.312264 & 0.997247 \\
			{LS2} & 14.8497 & 0.215213 & 0.983678 \\
			{Pd$_{77.5}$Cu$_{6}$Si$_{16.5}$} &  0.789078 & 0.0876754 & 0.998759 \\
			{Salol} & 17.1643 & 0.553687 & 0.993136 \\
			{Diopside} & 13.1776  & 0.0941259 & 0.997362 \\
			{Anorthite} & 2.25807 & 0.141129 & 0.991396 \\
			{BS2} & 4.902 & 0.0505361 & 0.998646 \\
			{Albite} & 13.3105 & 0.511942 & 0.87503 \\
			Zr$_{74}$Rh$_{26}$ & 0.115181 & 0.000984452 & 0.983959 \\
			Pd$_{40}$Ni$_{40}$P$_{20}$ & 12.3782 & 0.515757 & 0.993153 \\
			Zr$_{57}$Ni$_{43}$ & 0.351164 & 0.00172139 & 0.977947 \\
			Cu$_{64}$Zr$_{36}$ & 0.190441 & 0.00307162 & 0.984655 \\
			Ni$_{34}$Zr$_{66}$  & 0.121782 & 0.00162376 & 0.993343 \\
			Zr$_{50}$Cu$_{48}$Al$_{2}$ & 10.617 & 0.312264 & 0.997247 \\
			Ni$_{62}$Nb$_{38}$ & 0.448888 & 0.00487922 & 0.9841 \\
			Vit106a & 6.23195 & 0.623195 & 0.996508 \\
			Cu$_{55}$Zr$_{45}$ & 0.223386 & 0.00314628 & 0.987581 \\
			H$_2$O & 0.00731595 & 0.000215175 & 0.999412 \\
			Glucose & 1.48859 & 0.0513308 & 0.999499 \\
			Glycerol & 76.0137 & 1.85399 & 0.945217 \\
			Ti$_{40}$Zr$_{10}$Cu$_{30}$Pd$_{20}$ & 0.395717 & 0.00316573 & 0.988712 \\
			Zr$_{70}$Pd$_{30}$ & 0.080497 & 0.00134162 & 0.996159 \\
			Zr$_{80}$Pt$_{20}$ & 0.077876 & 0.00162242 & 0.971562 \\
			NS2 & 20.9749 & 0.723273 & 0.981462 \\
			Cu$_{60}$Zr$_{20}$Ti$_{20}$ & 0.196626 & 0.0012063 & 0.985095 \\
			Cu$_{69}$Zr$_{31}$ & 0.756104 & 0.00804366 & 0.950419 \\
			Cu$_{46}$Zr$_{54}$ & 0.650675 & 0.00971157 & 0.910136 \\
			Ni$_{24}$Zr$_{76}$ & 0.0453595 & 0.0008584 & 0.991683 \\
			Cu$_{50}$Zr$_{42.5}$Ti$_{7.5}$ & 0.0535541 & 0.00172755 & 0.982531 \\
			D Fructose & 0.554086 & 0.0240907 & 0.946689 \\
			TNB1 & 8.97792 & 0.448896 & 0.996155 \\
			Selenium & 6.43906 & 0.292684 & 0.995906 \\
			CN60.40 & 0.746426 & 0.0678569 & 0.998937 \\
			CN60.20 & 0.147407 & 0.0105291 & 0.999883 \\
			Pd$_{82}$Si$_{18}$ & 1.2915 & 0.1435 & 0.998916 \\
			Cu$_{50}$Zr$_{45}$Al$_{5}$ & 0.109111 & 0.000742252 & 0.992842 \\
			Ti$_{40}$Zr$_{10}$Cu$_{36}$Pd$_{14}$ & 0.195736 & 0.00163113 & 0.92674 \\
			Cu$_{50}$Zr$_{50}$ & 0.235607 & 0.00420727 & 0.976969 \\
			Isopropyl benzene & 4.47953 & 0.344579 & 0.993307 \\
			Butylbenzene & 1.97384 & 0.140989 & 0.995543 \\
			Cu$_{58}$Zr$_{42}$ & 0.551631 & 0.0108163 & 0.966384 \\
			Vit 1 & 46.5891 & 2.58828 & 0.956556 \\
			Trehalose & 8.93373 & 0.288185 & 0.934837 \\
			Sec-Butylbenzene & 1.27723 & 0.159653 & 0.976809 \\
			SiO$_2$ & 57.7053 & 1.98984 & 0.660326 \\
			\botrule
		\end{tabular}}
	\end{table*}

\section{Conclusion}
We advanced a classical statistical mechanical framework for understanding the dynamics of supercooled liquids. We demonstrated, both qualitatively and quantitatively, that the resultant expression  that is predicted by this classical approach (and by an earlier companion quantum version \cite{bib:DEH1}) for the viscosity of supercooled liquids below the melting temperature can describe/reproduce the behavior of all liquids studied to objectively high accuracy. We demonstrated that the viscosity data of 45 different liquids can be collapsed onto a single scaling curve, suggesting that an underlying universality may be present in the dynamical behavior of supercooled liquids. Further support on our results appears in \cite{bib:DEH2}. We hope that our newly found universal 16 decade collapse for the viscosity data of all known liquid types and the theoretical ideas that led us to it will prompt further discussion on the underlying phenomenology of supercooled liquids and the glass transition.

\section{Acknowledgements}
NW and ZN were supported by the NSF DMR-1411229. CP and KFK were supported by the NSF DMR 12-06707, NSF DMR 15-06553, and NASA-NNX 10AU19G. ZN thanks the Feinberg foundation visiting faculty program at Weizmann Institute. The initial version of this work was completed while ZN was at the Aspen Center for
Physics, which is supported by the National Science Foundation grant PHY-1066293. NW would like to sincerely thank Robert Ashcraft and Rongrong Dai for supplying data and stimulating discussions. We further thank Hajime Tanaka and Takeshi Egami for their kind words and encouragement.

\end{document}